\documentclass[matbio, referee]{svjour}

\usepackage{graphicx}
\usepackage{amsmath}

\def\A{\mathcal{A}}
\def\S{\mathcal{S}}
\def\I{\mathcal{I}}
\def\R{\mathcal{R}}
\def\J{\mathcal{J}}
\def\pr{\mathrm{Pr}}

\title{SIR dynamics in random networks with heterogeneous connectivity}
\titlerunning{SIR dyn. in rand. net. w. heter. conn.}
\author{Erik Volz}

\institute{{Department of Integrative Biology, University of Texas, Austin, TX,~\email{erik.volz@mail.utexas.edu}}}

\date{Received: January 17, 2007}

\begin{document}
\maketitle

\abstract{
Random networks with specified degree distributions have been proposed as realistic models of population structure, yet the problem of dynamically modeling SIR-type epidemics in random networks remains complex. I resolve this dilemma by showing how the SIR dynamics can be modeled with a system of three nonlinear ODE's. The method makes use of the probability generating function (PGF) formalism for representing the degree distribution of a random network and makes use of network-centric quantities such as the number of edges in a well-defined category rather than node-centric quantities such as the number of infecteds or susceptibles. The PGF provides a simple means of translating between network and node-centric variables and determining the epidemic incidence at any time. The theory also provides a simple means of tracking the evolution of the degree distribution among susceptibles or infecteds. The equations are used to demonstrate the dramatic effects that the degree distribution plays on the final size of an epidemic as well as the speed with which it spreads through the population. Power law degree distributions are observed to generate an almost immediate expansion phase yet have a smaller final size compared to homogeneous degree distributions such as the Poisson. The equations are compared to stochastic simulations, which show good agreement with the theory. Finally, the dynamic equations provide an alternative way of determining the epidemic threshold where large-scale epidemics are expected to occur, and below which epidemic behavior is limited to finite-sized outbreaks. \\
}

\section{Introduction\label{sec:intr}}

Contact patterns constitute an important aspect of heterogeneity within a population of susceptible and infectious individuals, but it has been a difficult factor to incorporate into epidemiological models. Compartment models can capture many aspects of population heterogeneity, such as with respect to heterogeneous susceptibility and infectiousness~\cite{veli1,and,die}, however such models usually assume individuals mix homogeneously within each category. In contrast, the contact patterns responsible for the spread of many infectious diseases tend to be characterized by constant relationships marked by high levels of heterogeneity in the number of contacts per individual.

An alternative approach is to model a population of susceptibles and infecteds and the contact patterns among them as a static random network ~\cite{lilj1,strog1,newm2,andeMay2}. This approach has generated a new category of epidemiological models in which epidemics spread from node to node by traversing network connections~\cite{satoVesp1,meyePourNewmSkowBrun1,newm1,warr1,dezsoBara1,saraKask1}. Random networks with specified degree distributions have been proposed as a simple but realistic models of population structure. This case has the advantage of being well understood mathematically. The expected final size of epidemics in random networks with a given degree distribution has been solved exactly~\cite{meyePourNewmSkowBrun1,newm1}. 
The network approach has the advantage that the mathematics of stochastic branching processes~\cite{wilf1,harr1,athrNey1} can be brought to bear on the problem. This allows for precise descriptions of the distribution of outbreak sizes early in the course of the epidemic as well as the final size.~\cite{meyePourNewmSkowBrun1,newm1} 

A shortcoming of the network model is that it is difficult to describe the explicit dynamical behavior of epidemics on networks. The distribution of outbreak sizes is easy to calculate, yet the dynamic epidemic incidence, that is the number of infecteds at a time $t$, has been difficult to derive. Simulation has been used in this case~\cite{euba1}. 

Heterogeneity in the number of contacts within networks makes it difficult to derive differential equations to describe the course of an epidemic.
Nevertheless, several researchers~\cite{barthBarrSatoVesp1,satoVesp2,satoVesp3,boguSatoVesp1,eameKeel1} have been successful modeling many of the dynamical aspects of network epidemics, particularly in the early stage where asymptotically correct equations for disease incidence are known. These solutions break down, however, when the finite size of a population becomes a significant factor. 
We improve upon these results by presenting a system of nonlinear ordinary differential equations which can be used to solve for epidemic incidence at any time, from an initial infected to the final size, as well as other quantities of interest. We treat the simplest possible case of the SIR dynamics with constant rate of infection and recovery. Section~\ref{sec:mode} describes the model. Several examples are given in section~\ref{sec:exam}, and section~\ref{sec:simu} compares the analytical results to stochastic simulations.

\section{SIR in Random Networks~\label{sec:mode}}
	
	The networks considered here are random networks with an arbitrary degree distribution $p_k$ ($p_k$ being the probability of a random node having degree $k$)~\cite{newm2,mollReed1}. Nodes can be in any of three exclusive states: susceptible ($\S$),  infectious ($\I$), or recovered ($\R$). The dynamics are as follows. When a node is infectious, it will transmit infection to each of its neighbors independently at a constant rate $r$. Infectious nodes  become recovered at a constant rate $\mu$, whereupon they will no longer infect any neighbors. This will be made precise in the next section. 
	
	It is desirable to determine the dynamics of the number of susceptibles and infecteds and to develop equations in terms of those quantities. This, however, turns out to be intractable due to heterogeneity in the number of contacts. The problem can be resolved by developing equations in terms of dynamic variables representing network-based quantities, for example, the number of connections to susceptible or infectious nodes at a time $t$. The network- and node-based quantities are defined in the next section. 
	
	To bridge the divide between connection- and node-based quantities, a mathematical device known as a probability generating function (PGF)~\cite{wilf1} is extremely useful.The PGF has many useful properties and is frequently used in probability theory and the theory of stochastic branching processes.  Given a discrete probability density $p_k$, the PGF is defined as the series:
	\begin{align}
		g(x) &= p_0 + p_1 x + p_2 x^2 + p_3 x^3 + \cdots
	\end{align}
	The variable $x$ in the generating function  serves only  as a place-holder.  To illustrate the utility of this device, consider the possibility that the probability of a node being infected, say $\lambda$, is compounded geometrically according the node's degree. Then, the probability of a degree $k$ node being susceptible is $(1-\lambda)^k$, that is, the probability of not being infected along any of $k$ connections. If the hazard is identical for all nodes,  the cumulative epidemic incidence (the fraction of nodes infectious or recovered) will be
	\begin{align}
		J &=1 - [p_0 (1-\lambda)^0  + p_1 (1-\lambda)^1 + p_2 (1-\lambda)^2 + \cdots ]\\
		&= 1 - g(1-\lambda)
	\end{align}
	
	Table~\ref{tab:parms} gives a summary of the parameters used in the model.

	\subsection{Definitions}
		\begin{table}
		\begin{center}
			\caption{Parameters and dynamic variables for the network SIR model.}
			\label{tab:parms}
		\begin{itemize}
			\item $r:=$  Force of infection. The constant rate at which infectious nodes infect a neighbor.
			\item $\mu:=$ Recovery rate. The constant rate at which infected nodes become recovered.
			\item $p_k:=$ The probability that a node will have degree $k$.  
			\item $g(x):=$ The probability generating function for the degree distribution $p_k$. 
			\item $S:=$ The fraction of nodes susceptible at time $t$.
			\item $I:=$ The fraction of nodes infectious at time $t$. 
			\item $R:=$ The fraction of nodes recovered at time $t$.
			\item $J=I+R$ The cumulative epidemic incidence at time $t$.
			\item $\A_{X}$ Set of arcs $(ego,alter)$ such that node $ego$ is in set $X$. 
			\item $M_{X}$ Fraction of arcs in set $\A_X$. 
			\item $\A_{XY}$ Set of arcs $(ego,alter)$ s.t. $ego\in X$ and $alter\in Y$.
			\item $M_{XY}$ Fraction of arcs in set $\A_{XY}$. 
		\end{itemize}
		\end{center}
		\end{table}
	An undirected network can be defined as a graph $\mathcal{G}=\{ V, \mathcal{E}\}$ consisting of a set of vertices $V$ corresponding to the nodes in the network, and a set of edges $\mathcal{E}$ with elements of unordered pairs of vertices, $\{a,b\}$ where $a,b \in V$. Two vertices $a,b$ are said to be \emph{neighbors} or \emph{neighboring each other} or simply \emph{connected} if there exists an edge $e = \{a,b\}\in \mathcal{E}$. For the purposes of this model, the terms ``vertex'' and ``node'' will often be used interchangeably.
	
	For the random networks considered here, the probability of being connected to a node is proportional to the degree of that node. Denote the degree of a node $v\in V$ as $d_v$. 
	Then given an edge $\{a,x\}\in\mathcal{E}$, the probability that $x=b$ is $d_b/\sum_{i\in V} d_i$. This definition allows multiple edges to the same node as well as loops from a node to itself, however the existence of multiple edges and loops is exceedingly rare for large sparse random networks such that results based on this case can be safely applied to networks without multiple edges. Networks of this type can be generated by a variation\footnote{Note that this version of the configuration model allows loops and multiple-edges.} of the ``configuration model''~\cite{moRe95}:
	\begin{enumerate}
		\item To each node $v\in V$ assign an i.i.d. degree $\delta_v$ from distribution $p_k$
		\item Generate a new set $X$ of ``half-edges'' with $\delta_v$ copies of node $v$ for all nodes
		\item Insure $X$ has an even number of elements, for example, by deleting a uniform random element if odd. 
		\item While $X$ is not empty, draw two elements $v_1, v_2$ uniformly at random and create edge $\{v_1,v_2\}$. 
	\end{enumerate}
	
	At any point in time, a vertex can be classified as susceptible, infectious, or recovered. Let $\S,\I$, and $\R$ denote the disjoint sets of vertices classified as susceptible, infectious, or recovered respectively. $\J=\I\cup \R$ will denote the set of infectious or recovered nodes. $S,I,$ and $R$ will denote the fraction of nodes in the sets $\S,\I$, and $\R$ respectively. The cumulative epidemic incidence will be the fraction of nodes in set $\J$.
	
	As stated in the previous section, infectious vertices $a\in \I$ will infect neighboring susceptible vertices $b\in \S$ at a constant rate $r$. Infectious vertices will become recovered (move to set $\R$) at a constant rate $\mu$.
	
	Although the network is undirected in the sense that any two neighboring vertices can transmit infection to one another, we wish to keep track of who infects who. Therefore, for each edge $\{a,b\}\in \mathcal{E}$, let there be two arcs, which will be defined to be the ordered pairs $(a,b)$ and $(b,a)$. Let $\A$ denote the set of all arcs in the network. The first element in the ordered pair $(a,b)$ will frequently be called the \emph{ego} and the second element the \emph{alter}.  
	
	$\A_{XY}$ will denote the subset of arcs such that $ego\in X$ and $alter \in Y$. $\A_X$ will denote the subset of arcs such that $ego\in X$. $M_{XY} = \#\{\A_{XY}\}/\#\{\A\}$ will denote the fraction of arcs in the corresponding set $\A_{XY}$.
			
	For example, two variables will be especially important in the derivations that follow. $M_{SS}$ is the fraction of arcs with a susceptible ego and a susceptible alter. $M_{SI}$ is the fraction of arcs with a susceptible ego and and infectious alter. $M_S$ will be the fraction of arcs with a susceptible ego and an alter of any type.  
	
\subsection{Dynamics}
	
	Our objective is to develop a deterministic model to describe epidemic dynamics expressed with a low-dimensional system of differential equations. At first, this goal may seem incompatible with network-SIR dynamics described in the last section. Infection spreads along links in a random network, which implies the epidemic incidence at any time as well as the final size must also be random, depending on the particular structure of a given random network. This is true, however it is possible to avoid such considerations by focusing on epidemic dynamics in the limit as population size goes to infinity. This strategy has been used in previous work to calculate the expected final size of epidemics in infinite random networks~\cite{newm1} expressed as a fraction of the total population size. A similar strategy is followed here by considering the fraction of nodes in sets $\S,\I,$ and $\R$, after a small fraction $\epsilon$ nodes are infected initially in a susceptible population. The conclusion is the system of equations given in table~\ref{tab:summary} in terms of the dynamic variables given in table~\ref{tab:dynVar}. The dynamics predicted by these equations are compared to stochastic simulations with large but finite networks in section~\ref{sec:simu}.

		\begin{table}
		\begin{center}
			\caption{Network-based dynamic variables for the network SIR model.}
			\label{tab:dynVar}
		\begin{itemize}
			\item $\theta:=$ The fraction of degree one nodes that remain susceptible at time $t$. 
			\item $p_I:= M_{SI}/M_S$.  The probability that an arc with a susceptible ego has an infectious alter. 
			\item $p_S:= M_{SS}/M_S$.  The probability that an arc with a susceptible ego has a susceptible alter.
		\end{itemize}
		\end{center}
		\end{table}

	Consider a susceptible node $ego$ at time $t$ with a degree $k$. Then there will be a set of $k$ arcs $\{ (ego, alter_1), (ego,alter_2), \cdots, (ego,alter_k)\}$ corresponding to $ego$.  We will assume that for each arc $(ego,alter_i)$ there will be a uniform probability $p_I = M_{SI}/M_S$ that $alter_i$ is infectious.  
	Then there is an expected fraction $k p_I$ arcs $(ego,alter)$ such that $alter$ is infectious. In a time $dt$, an expected number $r k p_I ~dt$ of these will be such that the infectious alter transmits to $ego$. Consequently, the hazard for ego becoming infected at time $t$ is
	\begin{equation}
		\label{eqn:lambda1}
		\lambda_k(t) = r k p_I(t)
	\end{equation}
	
	Now let $u_k(t)$ represent the fraction of degree $k$ nodes that remain susceptible at time $t$, or equivalently the probability that $ego$ in the previous example is susceptible. Using equation~\ref{eqn:lambda1}, 
	\begin{equation}
		\label{eqn:uk1}
		\begin{split}
		\displaystyle 	u_k(t) = \exp \{-\int_{\tau=0}^t \lambda_k(\tau) d \tau \}  = \exp \{-\int_{\tau=0}^t r k p_I(\tau) d \tau \}  \\
		= \exp \{-\int_{\tau=0}^t r  p_I(\tau) d \tau \}^k 
	 	\end{split}
	\end{equation}
	
	Subsequently we will use the symbol $\theta$ to denote $u_1 = \exp \{-\int_{\tau=0}^t r  p_I(\tau) d \tau \}$. From equation~\ref{eqn:uk1} it is clear that $u_k = \theta^k$. 
	
	Given $\theta$, it is easy to determine the fraction of nodes which remain susceptible at a time $t$.
	\begin{equation}
		\label{eqn:S1}
		\begin{split}
		S = p_0 + p_1 u_1 + p_2 u_2 + p_3 u_3 \cdots \\
		= p_1 \theta + p_2 \theta^2  + p_3 \theta^3 + \cdots = g(\theta)  \\
		\end{split}
	\end{equation}
	This equation makes use of the generating function $g(\cdot)$ for the degree distribution which greatly simplifies this and subsequent equations. 
	
	The dynamics of $\theta$ are dependent on the hazard $\lambda_1$.
	\begin{equation}
	\label{eqn:thetadot}
	\begin{split}
	 {\displaystyle \frac{d \theta/dt}{\theta} = -\lambda_1(t)  } \Rightarrow\\
		\dot{\theta} = -\theta \lambda(t) = -\theta ~r ~ p_I
	\end{split}
	\end{equation}
	
	Unfortunately, this does not completely specify the dynamics of $\theta$ and by extension $S$, which also depends on the variable $p_I$. The derivation of the dynamics of $p_I$ follows. 
	\begin{equation}
		\label{eqn:dotpi1}
	\displaystyle \dot{p}_I = \frac{d}{dt} \frac{M_{SI}}{M_S} = \frac{\dot{M}_{SI}}{M_S}-\frac{\dot{M}_S M_{SI}}{M_{S}^2}
	\end{equation}
	
	Our goal is to put equation~\ref{eqn:dotpi1} in terms of the variables $\theta,p_S,p_I$ and the PGF $g(\cdot)$. $M_S$ is easily placed in terms of these variables. 
	\begin{equation}
	\label{eqn:ms}
	\begin{split}
	M_S =  \sum_k p_k \times k \times \pr[\mathrm{degree~k~node~susceptible}] / \sum_k k p_k\\
	 {\displaystyle = \sum_k p_k k \theta^k / g'(1) = \left[ \frac{d}{dx} g(\theta x) \right]_{x=1} / g'(1) = \theta g'(\theta)} / g'(1) \\
	 \end{split}
	\end{equation}
	
	$M_{SI}$ follows easily.
	\begin{equation}
	\label{eqn:msi}
	M_{SI} = M_S \times M_{SI} / M_S = M_S p_I = p_I \theta g'(\theta) / g'(1)
	\end{equation}
	
	In time $dt$, $-\dot{S}$ nodes become infectious. Since $S=g(\theta)$, 
	\begin{equation}
		\label{eqn:dots}
		\dot{S} = \frac{d}{dt} S = \frac{d}{dt} g(\theta) = \dot{\theta} g'(\theta) = - r p_I \theta  g'(\theta)
	\end{equation}
	
	Calculating $\dot{M}_{SI}$ requires careful consideration of the rearrangement of arcs among sets $\A_{SS}$ and $\A_{SI}$ as $-\dot{S}$ nodes become infected in a small time interval. Since the hazard of becoming infected is proportional to the number of arcs to an infectious alter, a newly infected node will be selected with probability proportional to the number of arcs from the node to infectious nodes. 
	
	To clarify subsequent calculations, I will introduce the notation $\delta_{XY}$ to represent the average degree of nodes in set $X$, selected with probability proportional to the number of arcs to nodes in set $Y$, not counting one arc to nodes of type $Y$. For example, if we select an arc $(ego\in X, alter\in Y)$ uniformly at random out of the set of arcs from nodes in set $X$ to nodes in set $Y$ ($\A_{XY}$), and follow it to the node in set $X$, ($ego$), then $\delta_{XY}$ will represent the average number of arcs $(ego, alter')$ not counting the arc we followed to $ego$. This is commonly called the ``excess degree'' of a node~\cite{meyers2005nta}. Furthermore, $\delta_{XY}(Z)$ will be as $\delta_{XY}$ but counting only arcs from $ego$ to nodes in set $Z$,  $(ego,alter\in Z)$.  
	
	To calculate $\dot{M}_{SI}$ we need to first calculate $\delta_{SI}$, and for this it is necessary to derive the degree distribution among susceptible nodes. It is necessary to assume\footnote{
		Although a rigorous proof for this is currently lacking, it is borne out by the success of this mathematical theory in predicting epidemic final size and dynamics (see sections~\ref{sec:exam} and~\ref{sec:simu} below).
	}
	that arcs from a susceptible ego to nodes in sets $\S,\I,\R$ are distributed multinomially with probabilities $p_S,p_I,$ and $p_R=1-p_S-p_I$ respectively. Let $d_{ego}(X)$ be the r.v. denoting the number of arcs from $ego$ to nodes in set $X$. 
	Letting $c$ normalize the distribution, and letting the dummy variables $x_S,x_I,$ and $x_R$ correspond to the number of arcs from a susceptible ego to an alter in sets $\S,\I,\R$ respectively, the degree distribution for susceptible nodes will be generated by
	\begin{equation}
	\label{eqn:pks0}
	\begin{split}
	g_S(x_S, x_I, x_R) = \sum_k p_k u_k \sum_{i,j|i+j\leq k} x_S^i x_I^j x_R^{k-i-j} \pr[d(S)=i,d(I)=j|p_S,p_I]  / c\\
	\end{split}
	\end{equation}
	Using the multinomial theorem this becomes
	\begin{equation}
	\label{eqn:pks}
	\begin{split}
	g_S(x_S, x_I, x_R) = \sum_k p_k \theta^k (x_S p_S + x_I p_I + x_R (1-p_S-p_I))^k / c  \\
	 = g(\theta(x_S p_S + x_I p_I + x_R (1-p_S-p_I))) / g(\theta),
	\end{split}
	\end{equation}
	where $c=\sum_k p_k \theta^k (p_S +  p_I + (1-p_S-p_I))^k = g(\theta)$ normalizes the distribution. 
	
	The degree distribution for susceptible nodes selected with probability proportional to the number of arcs to infectious nodes will be generated by the following equation. Note that this equation \emph{does not} count one arc to infectious nodes.  
	\begin{equation}
	\label{eqn:pksi}
	\begin{split}
	g_{SI}(x_S,x_I,x_R)  = \\
	\sum_k p_k u_k \sum_{i,j|i+j\leq k}j\times x_S^i x_I^j x_R^{k-i-j} \pr[d(S)=i,d(I)=j|p_S,p_I]  /\\
	\sum_k p_k u_k \sum_{i,j|i+j\leq k} j\times\pr[d(S)=i,d(I)=j|p_S,p_I] \\
	= \left[ \frac{d}{d x_I} g_{S}(x_S, x_I, x_R)  \right] / \left[ \frac{d}{d x_I} g_{S}(x_S,x_I,x_R)  \right]_{x_S=x_I=x_R=1}\\
	= g'(\theta(x_S p_S + x_I p_I + x_R (1-p_S-p_I))) / g'(\theta)  
	\end{split}
	\end{equation}
	Because arcs are distributed multinomially to nodes in sets $\S,\I,\R$, we have $g_{SS}(x_S,x_I,x_R) = g_{SI}(x_S,x_I,x_R)$, which is easy to verify by repeating the calculation in equation~\ref{eqn:pksi}. 
	
	A useful property of PGF's is that the mean of the distribution they generate can be calculated by differentiating and evaluating with the dummy variables set to one~\cite{wilf1}. Now using equations~\ref{eqn:pks} and ~\ref{eqn:pksi}, we have the following results. 
	\begin{eqnarray}
		\label{eqn:deltasi}
		\delta_{SI} = \left[ \frac{d}{d x} g_{SI} (x,x,x) \right]_{x=1} = \theta g''(\theta)/g'(\theta) \\
		\label{eqn:deltasii}
		\delta_{SI}(I) = \left[ \frac{d}{d x_I} g_{SI} (x_S,x_I,x_R) \right]_{x_S=x_I=x_R=1} =  p_I \theta g''(\theta) / g'(\theta)  \\
		\label{eqn:deltasis}
		\delta_{SI}(S) = \left[ \frac{d}{d x_S} g_{SI} (x_S,x_I,x_R) \right]_{x_S=x_I=x_R=1} = p_S \theta g''(\theta) / g'(\theta)
	\end{eqnarray}

	As a fraction $-\dot{S}$ nodes leave set $\S$ in time $dt$, the fraction of arcs between $\S$ and $\I$, $M_{SI}$ is reduced by the fraction of arcs from infectious nodes to the $-\dot{S}$ newly infectious nodes. Therefore $M_{SI}$ is reduced at rate $-\dot{S} \delta_{SI}(I)/g'(1)$. Because $\delta_{SI}(I)$ does not count the arc along which a node was infected, $M_{SI}$ is also reduced at a rate $r M_{SI}$ to account for all arcs which have an infectious ego which transmits to the susceptible alter. And in time $d t$, $\mu I$ nodes become recovered. The average number of arcs in $\A_{IS}$ per infectious node is proportional to $M_{SI}/I$. 
	Then $M_{SI}$ is reduced at a rate $\mu I (M_{SI} / I)  = \mu M_{SI}$. 
	
	The quantity $M_{SI}$ is also increased, as new infected nodes have links to susceptible nodes. A newly infectious node will have on average $\delta_{SI}(S)$ arcs to susceptible nodes, so $M_{SI}$ is increased at a rate $-\dot{S} \delta_{SI}(S)/g'(1)$. 
	
	To summarize, $M_{SI}$ decreases at the sum of rates
	\begin{itemize}
		\item $-\dot{S} \delta_{SI}(I)/g'(1)$
		\item $r M_{SI}$
		\item $\mu M_{SI}$
	\end{itemize}
	And $M_{SI}$ increases at the sum of rates
	\begin{itemize}
		\item $-\dot{S} \delta_{SI}(S)/g'(1)$
	\end{itemize}
	Then applying equations~\ref{eqn:deltasii},~\ref{eqn:deltasis}, and~\ref{eqn:dots} we have
	\begin{equation}
		\label{eqn:dotmsi}
		\begin{split}
		\dot{M}_{SI} = ((-\dot{S}) \delta_{SI}(S) - (-\dot{S})  \delta_{SI}(I))/g'(1) - (r+\mu) M_{SI} \\
		= r p_I (p_S - p_I) \theta^2 g''(\theta)/g'(1) - (r+\mu) M_{SI}
		\end{split}
	\end{equation}
	
	Finally, it is necessary to determine the time derivative of $M_S$.
	\begin{equation}
		\label{eqn:dotms}
		\begin{split}
		\dot{M}_S = \frac{d}{dt} \theta g'(\theta)/g'(1) = (\dot{\theta} g'(\theta) + \theta \dot{\theta} g''(\theta))/g'(1) \\
		= (-r p_I \theta g'(\theta) - r p_I \theta^2 g''(\theta))/g'(1)
		\end{split}
	\end{equation}
	
	Now applying equations\footnote{The normalizing constant $g'(1)$ cancels out and could have been left out these equations.}~\ref{eqn:ms},~\ref{eqn:dotmsi}, and~\ref{eqn:dotms} to equation~\ref{eqn:dotpi1} we solve for $\dot{p}_I$ in terms of the PGF and $\theta$. 
	\begin{equation}
		\label{eqn:dotpi2}
		\displaystyle \dot{p}_I = r p_I p_S \theta \frac{g''(\theta)}{g'(\theta)} - p_I (1-p_I) r - p_I \mu
	\end{equation}
	
	This equation makes use of the variable $p_{S}$ which changes in time. Deriving the dynamics of this variable will complete the model. This calculation is very similar to that for $\dot{p}_{I}$.
	\begin{equation}
	\label{eqn:dotps1}
		\displaystyle \dot{p}_S =  \frac{d}{dt} \frac{M_{SS}}{M_S} = \frac{\dot{M}_{SS}}{M_S}-\frac{\dot{M}_S M_{SS}}{M_{S}^2}
	\end{equation}
	
	The calculation for $\dot{M}_{SS}$ is very similar to that for $\dot{M}_{SI}$. Newly infected nodes have on average $\delta_{SI}(S)$ arcs to other susceptibles, so that 
	\begin{equation}
		\label{eqn:dotmss}
		\begin{split}
		\dot{M}_{SS} = -2 \times (-\dot{S}) \delta_{SI}(S) / g'(1) \\
		  = -2 r p_I p_S \theta^2 g''(\theta) / g'(1)
		\end{split}
	\end{equation}
	where the factor of $2\times$ accounts for two arcs per edge. 
	
	Now applying equations~\ref{eqn:ms},~\ref{eqn:dotms},and~\ref{eqn:dotmss} to equation~\ref{eqn:dotps1}, we have
	\begin{equation}
		\label{eqn:dotmps2}
		\displaystyle \dot{p}_S = r p_I p_S \left( 1 - \theta \frac{g''(\theta)}{g'(\theta)} \right) 
	\end{equation}
	
	The complete system of equations is summarized in table~\ref{tab:summary}. 
	
	The fraction of infectious nodes can be solved by introducing a fourth dynamic variable. The infectious class increases at a rate $-\dot{S}$ and decreases at a rate $\mu I$. Therefore
	\begin{equation}
		\label{eqn:doti}
		\dot{I} = -r p_I \theta g'(\theta) - \mu I
	\end{equation}

		\begin{table}
		\begin{center}
			\caption{A summary of the nonlinear differential equations used to the describe the spread of a simple SIR type epidemic through a random network. The degree distribution of the network is generated by $g(x)$. }
			\label{tab:summary}
			\begin{tabular}{l}
			\\
			\hline
			$ {\displaystyle  \dot{\theta} = - r p_I \theta }    $\\
			${\displaystyle \dot{p}_I = r p_S p_I \theta \frac{g''(\theta)}{g'(\theta)}  - r p_I (1-p_I) - p_I \mu }$  \\
			${\displaystyle \dot{p}_S =  r p_S p_I \left( 1-\theta \frac{g''(\theta)}{g'(\theta)} \right)  }$ \\
			\hline
			$ S = g(\theta)$\\
			$\dot{I} = r p_I \theta g'(\theta) - \mu I$\\
			\hline
			\end{tabular}
		\end{center}
		\end{table}
		
	An advantage of dynamic modeling of epidemics in networks is that the time-evolution of variables besides incidence can be calculated. Above it was shown how to calculate the degree distribution among susceptible nodes (eqn.~\ref{eqn:pks}). Additionally, the degree distribution among nodes which are either infectious or recovered (set $\J$) can be calculated by taking the complement.
	\begin{equation}
		\label{eqn:pkj}
		g_{J}(x) = (g(x) - g(\theta x))/ (1 - g(\theta))
	\end{equation}
	
\subsection{Initial Conditions}
If a small fraction $\epsilon$ of the nodes in the network are selected uniformly at random and initially infected, we can anticipate the following initial conditions. The fraction of arcs with infectious ego will also be $M_I=\epsilon$, and since $\epsilon$ is small, there is low chance of two initial infecteds being connected. Therefore $M_{SI}\approx M_I = \epsilon$. $\theta$, which can be interpreted as the fraction of degree one nodes remaining susceptible will be $1-\epsilon$. And $M_S = 1- M_{SI} = 1-\epsilon$ because there are initially no recovered nodes. And $M_{SS}=M_S-M_{SI}=1-2\epsilon$. To summarize, 
\begin{enumerate}
	\item $\theta(t=0) = 1-\epsilon$
	\item $p_I(t=0) = M_{SI} / M_S =  \epsilon/(1-\epsilon) $
	\item $p_S(t=0) = M_{SS} / M_S = (1-2\epsilon) / (1-\epsilon)$
\end{enumerate}

\subsection{Epidemic threshold \label{sec:epiThresh}}
	Epidemic dynamics can fall into one of two qualitatively different regimes. Below a threshold in the ratio $r/\mu$, the final size ($I_\infty$) is necessarily proportional to the fraction of initial infectious nodes: $I_\infty \propto \epsilon$. But above this threshold, epidemics occur, and necessarily occupy a fraction of the population even as $\epsilon\rightarrow 0$.
	
	As per equation~\ref{eqn:lambda1}, the number of new infections in a small time interval is proportional to $p_I$. This is in contrast to compartment models in which the number of new infections is proportional the current number of infectious. If $\dot{p}_I(t=0) < 0$, an epidemic will necessarily die out without reaching a fraction of the population. The epidemic threshold occurs where
	\begin{equation}
		\label{eqn:epiThresh1}
		\dot{p}_I(t=0) =0 = r p_S p_I \theta \frac{g''(\theta)}{g'(\theta)}  - r p_I (1-p_I) - p_I \mu 
	\end{equation}
	Applying the initial conditions given in the last section and considering $\epsilon \ll 1$ gives
	\begin{equation}
		\label{eqn:epiThresh2}
		\begin{split}
		{\displaystyle \dot{p}_I(t=0)= r \frac{1-2\epsilon}{1-\epsilon} \frac{\epsilon}{1-\epsilon} (1-\epsilon) g''(\theta)/g'(\theta) - r \frac{\epsilon}{1-\epsilon} \frac{1-2\epsilon}{1-\epsilon} - \mu \frac{\epsilon}{1-\epsilon}   } \\
		{\displaystyle = \epsilon \left( r \frac{g''(\theta)}{g'(\theta)} - r - \mu \right) = 0 }
		\end{split}
	\end{equation}
	Rearranging yields the critical ratio $r/\mu$ in terms of the PGF.
	\begin{equation}
		\label{eqn:epiThresh3}
		{\displaystyle (r/\mu)^* = \frac{g'(1)}{ g''(1) - g'(1)} }
	\end{equation}
	
	The epidemic threshold in equation~\ref{eqn:epiThresh3} can also be put in terms of the the transmissibility, which is the probability that an infectious ego will transmit infection to a given alter. Integrating over an exponentially distributed duration of infectiousness $T$, the transmissibility $\tau$ is calculated to be
	\begin{equation}
		\label{eqn:tau1}
		\begin{split}
		\tau = \int_{T=0}^\infty \pr[\mathrm{transmit~prior~to~T}] \times \pr[\mathrm{recover~at~T}] dT \\
		= \int_{T=0}^\infty (1-e^{-rT}) (\mu e^{-\mu T}) d T = \frac{r}{r+\mu}
		\end{split}
	\end{equation}
	Then rearranging equation~\ref{eqn:epiThresh3} yeilds the epidemic threshold in terms of $\tau$.
	\begin{equation}
		\label{eqn:epiThresh4}
		\tau^* = g'(1)/g''(1)
	\end{equation}
	This is consistent with previous results based on bond-percolation theory~\cite{newm1}.

\section{Examples\label{sec:exam}}
The model has been tested on several common degree distributions:
        \begin{itemize}
                \item Poisson: $p_{k} = \frac{z^{k} e^{-z}}{k!}$. This is generated by
                \begin{equation}
                        g(x) = e^{z (x - 1)}  \label{eqn:poisson}
                \end{equation}
                \item Power-law. For our experiments, we utilize power-laws with exponential cutoffs $\kappa$: $p_{k} = \frac{k^{-\gamma}e^{-k/\kappa}}{Li_{\gamma}(e^{-1/\kappa})}, k\geq 1$ where $Li_{n}(x)$ is the nth polylogarithm of x. This is generated by
		\begin{equation}
			g(x) = Li_\gamma(x e^{-1/\kappa}) / Li_\gamma(e^{-1/\kappa})  \label{eqn:powerlaw}
		\end{equation}
                \item Exponential: $p_{k} = (1-e^{-1/\lambda}) e^{-\lambda k}$. This is generated by
                \begin{equation}
                        g(x) = \frac{1 - e^{-1/\lambda}}{1 - x e^{-1/\lambda}} \label{eqn:exponential}
                \end{equation}
        \end{itemize}

\begin{figure}
		\begin{center}
			\includegraphics[width = .9\textwidth]{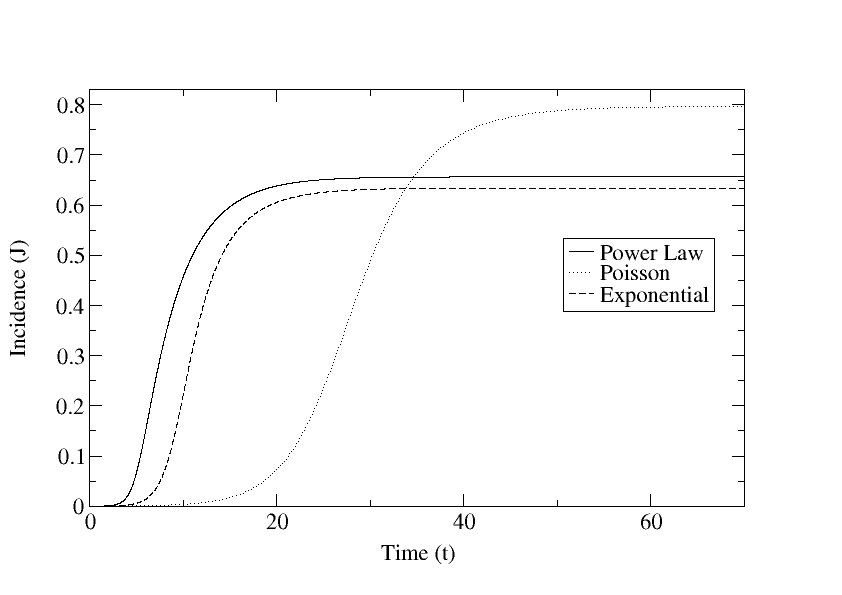}
			\caption{ The number of infecteds (including recovered) is shown versus time for an SIR model on three networks. Force of infection and mortality are constant: $r=0.2$, $\mu = 0.1$. The networks have Poisson ($z = 3$), power law ($\gamma = 1.615, \kappa = 20$), and exponential ($\lambda=3.475$) degree distributions. Each of these degree distributions has an average degree of 3.  }
			\label{fig:incidence}
		\end{center}
	\end{figure}

Figure~\ref{fig:incidence} shows the disease incidence for each of the degree distributions~(\ref{eqn:poisson}),~(\ref{eqn:powerlaw}), and~(\ref{eqn:exponential}), with a force of infection $r=.2$ and recovery rate $\mu=.1$. Initially $\epsilon = 10^{-4}$ nodes are infected. The parameters of the degree distributions were chosen so that each network has an identical average degree of 3.  That is, the density of connections in each network is the same. Nevertheless, there is widely different epidemic behavior due to the different degree distributions. 
Consistent with previous research, the degree distribution has a great impact on the final size of the epidemic~\cite{meyePourNewmSkowBrun1,newm1}. More importantly, the three networks exhibit widely varying dynamical behavior.
The power law network experiences epidemics which accelerate very rapidly. Such epidemics enter the expansion phase (the time at which incidence increases at its maximum rate) virtually as soon as the first individual in the network is infected. Both the Poisson and exponential networks experience a lag before the expansion phase of the epidemic. These observations are consistent with the findings in~\cite{barthBarrSatoVesp1} that the timescale of epidemics shortens with increasing contact heterogeneity. This has important implications for intervention strategies, as it is often the case that interventions are planned and implemented only after a pathogen has circulated in the population for some time. If an epidemic were to occur in the power law network, there would be little time to react before the the infection had reached a large proportion of the population.

		\begin{figure}
		\begin{center}
			\includegraphics[width = .9\textwidth]{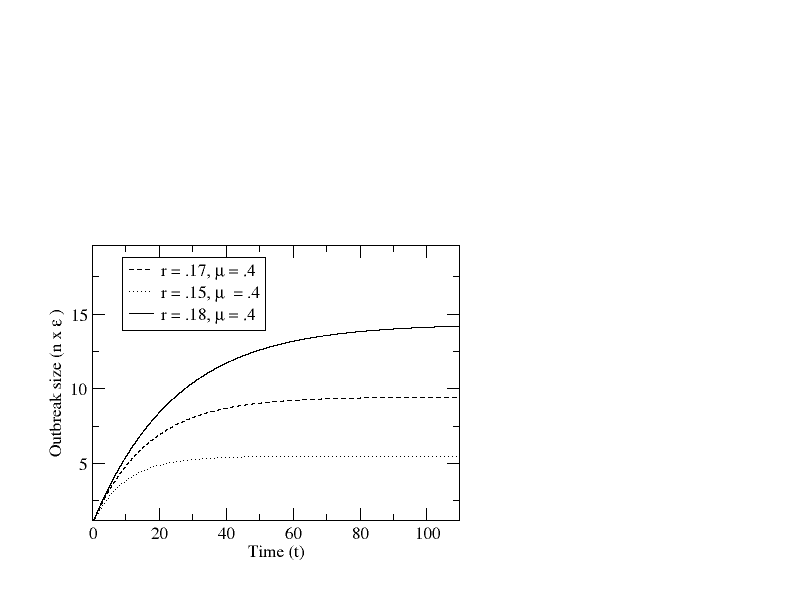}
			\caption{ The number of infecteds (including recovered) is shown versus time for an SIR model on a Poisson network ($z=3$). Each of these trials are below the epidemic threshold required to sustain an epidemic. The outbreak size is reported as a multiple of the fraction of initial infecteds in the network. Mortality is constant, $\mu = 0.4$, while three different levels of the force of infection are tried, $r=0.15,0.17,0.18$.}
			\label{fig:outbreak}
		\end{center}
	\end{figure}	
	
Recall from section~\ref{sec:epiThresh} that below the epidemic threshold $\tau^*$, only small, finite-sized outbreaks will occur. Figure~\ref{fig:outbreak} shows the qualitatively different dynamical behavior of outbreaks below the phase transition for networks with a Poisson distribution. Below the phase transition, the final size is always proportional to the fraction of initial infecteds $\epsilon$.

\begin{figure}
		\begin{center}
			\includegraphics[width = .9\textwidth]{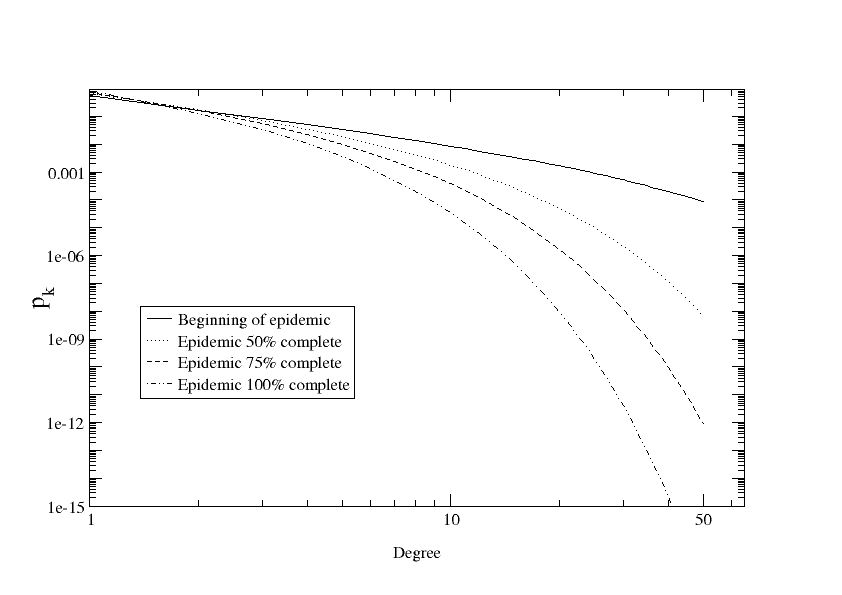}
			\caption{The degree distribution for susceptible nodes where the epidemic size is 50\%, 75\%, and 100\% of the final size, as well as degree distribution at the beginning of the epidemic. The degree distribution for the network as a whole is a power law with exponential cutoff (equation~\ref{eqn:powerlaw}).  }
			\label{fig:dist}
		\end{center}
	\end{figure}

Something offered by this model and not to the author's knowledge seen previously, is an explicit calculation for how the degree distribution of susceptibles evolves over the course of the epidemic. We expect the degree distribution to become bottom-heavy, as high degree nodes are gradually weeded out of the population of susceptibles. This is indeed observed in figure~\ref{fig:dist} for the Poisson trial described above. 

Recall that the degree distribution of susceptibles is generated by the multi-variate PGF~(\ref{eqn:pks}). The explicit degree distribution can be retrieved from equation~\ref{eqn:pks} by differentiation. The following gives the probability that a susceptible node has $m$ links at a time corresponding to $\theta$. 
\begin{equation}
{\displaystyle p_m^S = [ \frac{d^k}{dx^k} g_S(x,x,x) ]_{x=0}  / k!} 
\end{equation}
For example, applying this to the Poisson PGF (equation~(\ref{eqn:poisson})) gives
\begin{equation}
{\displaystyle   p_k = \frac{(z\theta)^ke^{-z\theta}}{k!}  } \label{eqn:dist2}
\end{equation}
which is simply the Poisson distribution with an adjusted parameter $z\times\theta$. Another example is illustrated in figure~\ref{fig:dist}, which shows the degree distribution among susceptibles for the power-law network considered above.
	
\subsection{Stochastic Simulations\label{sec:simu}}
	
		\begin{figure}
		\begin{center}
		\includegraphics[width=.8\textwidth]{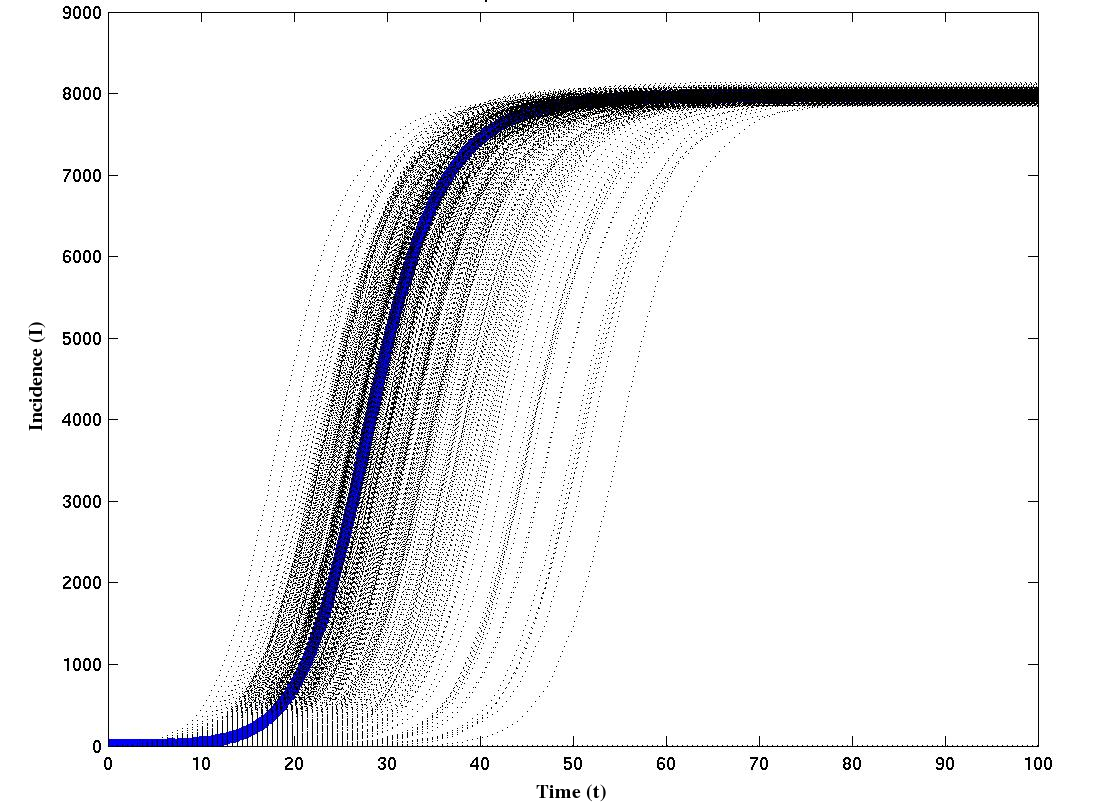}
		\caption{450 simulation trajectories of the cumulative epidemic incidence $J$ (dotted lines) for a Poisson ($z=3$) random network. The solid blue line shows the analytical solution.}
		\label{fig:simPois}
		\end{center}
		\end{figure}

	Simulation of SIR on networks presents two challenges: A random network must be generated with the desired degree distribution. Secondly, the stochastic rules that govern the transmission of disease at the microscopic scale must be well-defined, and an algorithm must be developed to aggregate this behavior into a large-scale simulation. 
	
	The random generation of networks with a given degree distribution is a well-explored problem. The first algorithm was proposed by Molloy and Reed~\cite{moRe95} which I have used for these experiments. Subsequent research has shown that imperfections can arise in the networks generated by this algorithm, but such biases should be tolerably small for these purposes~\cite{newmanalgorithm}.
	
	The simulation dynamics are as follows: 
	\begin{itemize}
	\item A node is chosen uniformly at random from the network as an initial infected.
	\item An infected node $v$ will recover after an exponentially distributed random time interval $\Delta t_\mu \sim Exp(\mu)$.
	\item When  a node $v$ is infected, each arc $(v,x)$ has a time of infection $\Delta t_x$  drawn from an exponential distribution $Exp(r)$. If $\Delta t_x < \Delta t_\mu$, node $x$ is infected after time $\Delta t_x$. Otherwise $x$ is not infected by $v$. 
	\end{itemize}
	This process continues until there are no more infectious nodes. 
	
	Figure~\ref{fig:simPois} shows the results of 450 simulations for the Poisson random network considered in the last section ($z=3$) with $10^4$ nodes. The black dotted line represents an independent simulation trajectory. The thick, blue line that cuts through the dense mass of simulation trajectories is the analytical trajectory based on the equations in table~\ref{tab:summary}. The initial conditions were chosen as in the previous section using $\epsilon = 10^{-4}$. 
	
\begin{figure}
	\begin{center}
	\includegraphics[width=.8\textwidth]{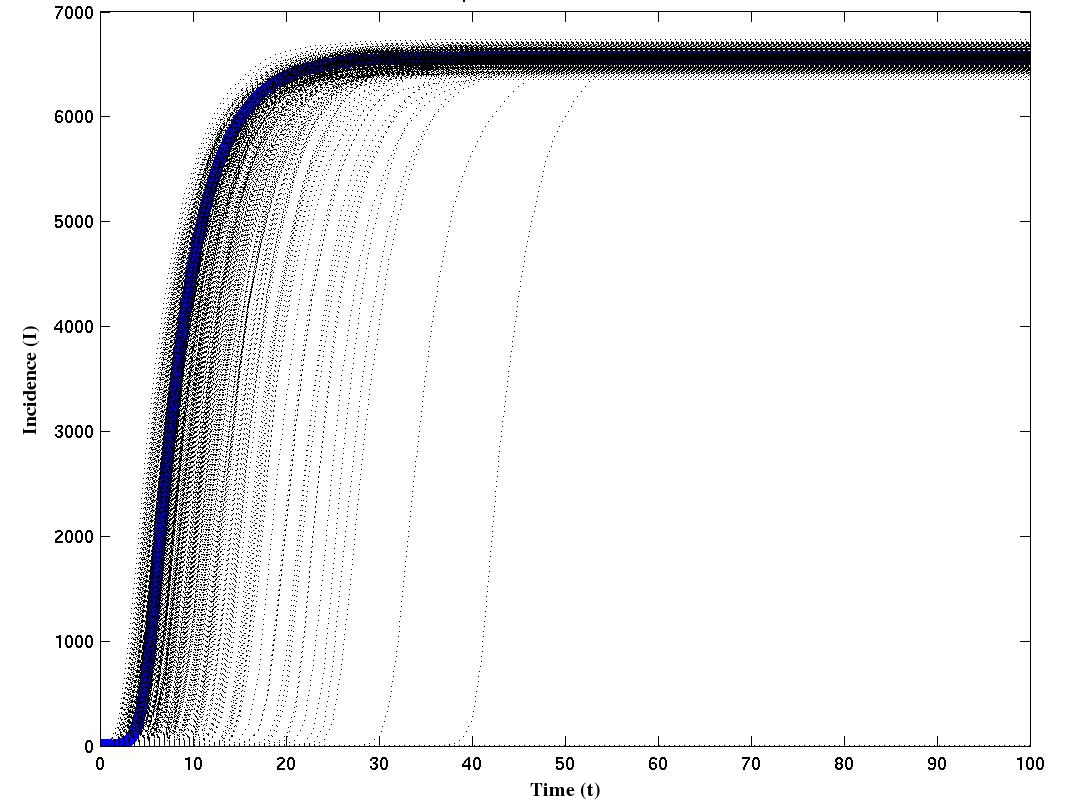}
	\caption{450 simulation trajectories of the cumulative epidemic incidence $J$ for a power law ($\gamma=1.615, \kappa=20$) random network. The solid line shows the analytical solution based on the system of equations in table~\ref{tab:summary}}
	\label{fig:simPl}
	\end{center}
	\end{figure}

	Figure~\ref{fig:simPl} shows a similar series of simulations for the power law degree distribution considered in the last section. In both cases, the analytical trajectory traverses the region with the highest density of simulation trajectories. The simulation trajectories also exhibit significant variability in the time required to reach the expansion phase and final size. This is largely due to the significant impact of random events early on in the epidemic. For example, an initial infected with a low average degree, or one which takes an inordinate amount of time to infect the next infected can markedly delay the onset of the expansion phase.

\begin{figure}
	\begin{center}
	\includegraphics[width=.8\textwidth]{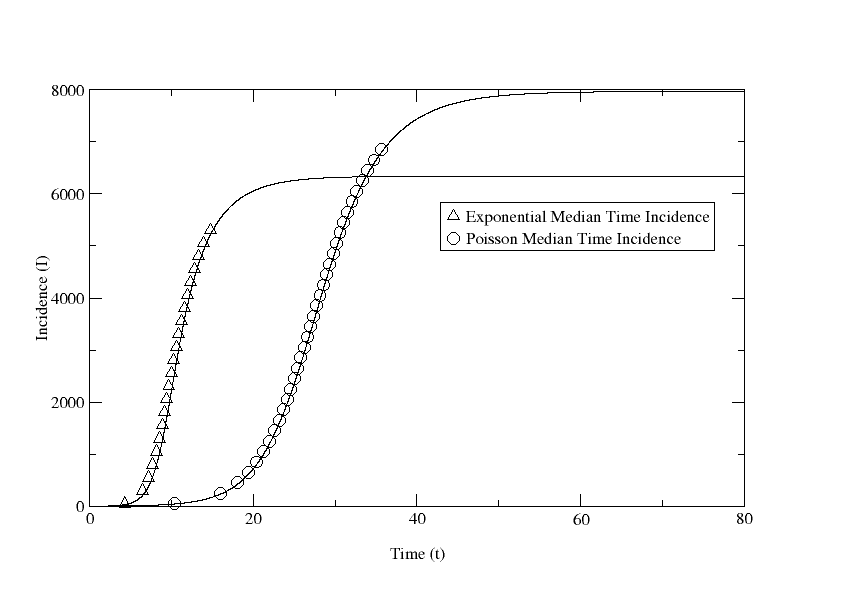}
	\caption{The median time required to reach a given incidence $J$ is shown for a Poisson network ($z=3$, circles) and an exponential network ($\lambda=3.475$, triangles). The solid line shows the analytical solution based on the system of equations in table~\ref{tab:summary}. }
	\label{fig:medT}
	\end{center}
	\end{figure}

	Figure ~\ref{fig:medT} shows the median-time incidence for the exponential and Poisson networks discussed in the last section. The data points show the median time required to reach a given incidence among 450 simulation trajectories. The solid line shows the analytical trajectory based on the system of equations given in table~\ref{tab:summary}.  Intuitively, the data points are showing the path of the most central trajectory from the swarm of simulation trajectories such as in figure~\ref{fig:simPois}.
	
\section{Discussion\label{sec:disc}}

The statistical properties of SIR epidemics in random networks have been understood for some time, but the explicit dynamics have been understood mainly through simulation. This paper has addressed this shortcoming by proposing a system of nonlinear ordinary differential equations to model SIR dynamics in random networks.  

It should be noted that the SI dynamics are a special case of this model ($\mu=0$), in which case the ultimate extent of the epidemic is simply the giant component\footnote{
The \emph{giant component} of a network, if it exists, is a set of nodes such there exists a path between any two of nodes, and furthermore occupies a non-zero fraction of the network in the limit as network size goes to infinity.}
 of the network.

The distribution of contacts, even holding the density of contacts constant, has enormous impact on epidemic behavior. This goes beyond merely the extent of the epidemic, but as shown here, the dynamical behavior of the epidemic. In particular, the distribution of contacts plays a key role in determining the onset of the expansion phase. 

The distribution dynamics from equation~\ref{eqn:pks} and shown in figure~\ref{fig:dist} have important implications for vaccination strategies. Previous work~\cite{kaplCrafWein1,hallLongNizaYang1} has focused on determining the critical levels of vaccination required to halt or prevent an epidemic. It is usually taken for granted that contact patterns among susceptibles are constant. Furthermore, most widespread vaccinations occur only once an epidemic is underway. Future research could be enhanced by considering optimal vaccination levels when the epidemic proceeds unhindered for variable amounts of time. 

It is hoped that the distribution dynamics described in this paper will find applications beyond modeling heterogeneous connectivity. The dynamic PGF approach might be used to capture other forms of heterogeneity, such as of susceptibility, mortality, and infectiousness.

\bibliographystyle{plain} 
\bibliography{volz-SIRNets-011707-2}

\begin{thebibliography}{10}

\bibitem{and}
R.M. Anderson and R.M. May.
\newblock {\em Infectious Diseases of Humans: Dynamics and Control}.
\newblock Oxford University Press, Oxford, 1991.

\bibitem{athrNey1}
K.~B. Athreya and P.~Ney.
\newblock {\em Branching Processes}.
\newblock Springer, New York, 1972.

\bibitem{barthBarrSatoVesp1}
M.~Barthelemy, A.~Barrat, R.~Pastor-Satorras, and A.~Vespignani.
\newblock Dynamical patterns of epidemic outbreaks in complex heterogeneous
  networks.
\newblock {\em J. of Theor. Biol.}, 235:275--288, 2005.

\bibitem{boguSatoVesp1}
M.~Boguna, R.~Pastor-Satorras, and A.~Vespignani.
\newblock Epidemic spreading in complex networks with degree correlations.
\newblock In J.M.~Rubi et. al., editor, {\em Statistical Mechanics of Complex
  Networks}, Berlin, 2003. Springer Verlag.

\bibitem{dezsoBara1}
Z.~Dezso and A.L. Barabasi.
\newblock Halting viruses in scale-free networks.
\newblock {\em Phys. Rev. E}, 65:055103(R), 2002.

\bibitem{die}
O.~Diekmann and J.A.P. Heesterbeek.
\newblock {\em Mathematical epidemiology of infectious diseases. Model
  building, analysis and interpretation}.
\newblock John Wiley \& Sons, Ltd., Chichester, 2000.

\bibitem{eameKeel1}
T.D. Eames and M.J. Keeling.
\newblock Modeling dynamic and network heterogeneities in the spread of
  sexually transmitted diseases.
\newblock {\em PNAS}, 99:13330--13335, 2002.

\bibitem{euba1}
S.~Eubank, H.~Guclu, V.S. Anil-Kunar, M.V. Marathe, A.~Srinivasan,
  Z.~Toroczkai, and N.~Wang.
\newblock Modelling disease outbreaks in realistic social networks.
\newblock {\em Nature}, 429:180--184, 2005.

\bibitem{andeMay2}
S.~Gupta, R.M. Anderson, and R.M. May.
\newblock Networks of sexual contacts: Implications for the pattern of spread
  of hiv.
\newblock {\em AIDS}, 3:807--817, 1989.

\bibitem{hallLongNizaYang1}
M.E. Halloran, I.~Longini, A.~Nizam, and Y.~Yang.
\newblock Containging bioterrorist smallpox.
\newblock {\em Science}, 298:1428, 2005.

\bibitem{harr1}
T.E. Harris.
\newblock {\em The Theory of Branching Processes}.
\newblock Springer, Berlin, 1963.

\bibitem{kaplCrafWein1}
E.H. Kaplan, D.L. Craft, and L.M. Wein.
\newblock Emergency response to a smallpox attach: The case for mass
  vaccination.
\newblock {\em PNAS U.S.A.}, 99:10935, 2002.

\bibitem{lilj1}
F.~Liljeros, C.R. Edling, L.A.N. Amaral, H.E. Stanley, and Y.~Aberg.
\newblock The web of human sexual contacts.
\newblock {\em Nature}, 411:907--908, 2001.

\bibitem{meyePourNewmSkowBrun1}
L.A. Meyers, B.~Pourbohloul, M.E.J. Newman, D.M. Skowronski, and R.C. Brun-ham.
\newblock Network theory and sars: Predicting outbreak diversity.
\newblock {\em J. Theor. Biol.}, 232:71--81, 2005.

\bibitem{meyers2005nta}
L.A. Meyers, B.~Pourbohloul, MEJ Newman, D.M. Skowronski, and R.C. Brunham.
\newblock {Network theory and SARS: predicting outbreak diversity}.
\newblock {\em J Theor Biol}, 232(1):71--81, 2005.

\bibitem{newmanalgorithm}
R.~Milo, N.~Kashtan, S.~Itzkovitz, M.~E.~J. Newman, and U.~Alon.
\newblock Uniform generation of random graphs with arbitrary degree sequences.
\newblock Preprint cond-mat/0312028, 2003.

\bibitem{moRe95}
Molloy and Reed.
\newblock A critical point for random graphs with a given degree sequence.
\newblock {\em Random Struct. and Algorithms}, 6:161, 1995.

\bibitem{mollReed1}
M.~Molloy and B.~Reed.
\newblock The size of the giant component of a random graph with a given degree
  sequence.
\newblock {\em Combinatorics, Probability and Computing}, 7:295--305, 1998.

\bibitem{newm1}
M.E.J. Newman.
\newblock The spread of epidemic disease on networks.
\newblock {\em Phys. Rev. E}, 66:016128, 2002.

\bibitem{newm2}
M.E.J. Newman, D.J. Watts, and S.H. Strogatz.
\newblock Random graph models of social networks.
\newblock {\em PNAS USA}, 99:2566--2572, 2002.

\bibitem{satoVesp2}
R.~Pastor-Satorras and A.~Vespignani.
\newblock Epidemic spreading in scale-free networks.
\newblock {\em Phys. Rev. Lett.}, 86:3200--3203, 2001b.

\bibitem{satoVesp3}
R.~Pastor-Satorras and A.~Vespignani.
\newblock Epidemic dynamics and endemic states in complex networks.
\newblock {\em Phys. Rev. E}, 63:066117, 2001c.

\bibitem{satoVesp1}
R.~Pastor-Satorras and A.~Vespignani.
\newblock {\em Handbook of Graphs and Networks: From the Genome to the
  Internet}, chapter Epidemics and immunization in scale-free networks.
\newblock Wiley-VCH, Berlin, 2002.

\bibitem{saraKask1}
J.~Saramki and K.~Kaski.
\newblock Modelling development of epidemics with dynamic small-world networks.
\newblock {\em J. Theor. Biol.}, 234:413--421, 2005.

\bibitem{strog1}
S.H. Strogatz.
\newblock Exploring complex networks.
\newblock {\em Nature}, 410:268--276, 2001.

\bibitem{veli1}
V.M. Veliov.
\newblock On the effect of population heterogeneity on dynamics of epidemic
  diseases.
\newblock {\em J. Math. Biol.}, 51:123--143, 2005.

\bibitem{warr1}
C.P. Warren, L.M. Sander, I.~Sokolov, C.~Simon, and J.~Koopman.
\newblock Percolation on disordered networks as a model for epidemics.
\newblock {\em Math. Biosci.}, 180:293--305, 2002.

\bibitem{wilf1}
H.S. Wilf.
\newblock {\em Generatingfunctionology}.
\newblock Academic Press, Boston, 2nd edition, 1994.

\end{thebibliography}

\end{document}